# Examining the Impact of Student Expectations on Undergraduate Biology Education Reform

Kristi L. Hall,[a] Jessica E. Watkins,[b] Janet E. Coffey,[a]
Todd J. Cooke,[c] & Edward F. Redish[b]

*Department of Curriculum & Instruction,[a] Department of Physics,[b] & Department of Cell Biology and Molecular Genetics,[c] University of Maryland, College Park MD 20742*

**Abstract.** The past 10-15 years have seen numerous calls for curricular reform in undergraduate biology education, most of which focus on changes to curriculum or pedagogy. Data collected from students in a large introductory undergraduate biology course indicate that student expectations about the nature of the knowledge they were learning influence how they interacted with reform efforts in that class. Given that student expectations influence the ways in which they participate in course activities, this paper (the first in a series that looks at student expectations in biology) argues that curriculum reform initiatives should consider student expectations in order to increase the chance for effective implementation.

## INTRODUCTION

The past 10-15 years have seen numerous calls for reforms in undergraduate biology education[1]. Traditionally, most introductory biology classes are taught in large (100+) student lectures. Professors often lecture with the aid of a PowerPoint presentation and classes generally follow the structure and topics outlined by the course textbook. Due to the size and structure of the lectures, there is rarely opportunity for interaction and dialogue between the students or between the students and the professor. However, most professors will set aside office hours to provide students an opportunity to get further clarification on lecture topics. There may also be graduate-student led review or discussion sections set up for this same purpose. Often no homework is assigned during the semester; therefore, 2-5 exams comprise almost all performance assessment for these courses. Due to the high volume of students, it is also common for the exams to be multiple-choice, scantron-style exams.

Reform-minded policy makers and educators claim this type of biology curriculum is outdated and ineffective. They assert that a large lecture with little or no interaction between students or students and professors emphasize the wrong types of skills (mainly rote memory) and do not teach the communication, reasoning, and analytic skills students need in order to thrive in an increasingly science and technology-driven world[2]. They stress that way we currently teach biology has not kept pace with the radical advances made in experimental biology and urge instructors to start teaching students about the science we do today, and the science we will be doing tomorrow, instead of continuing to teach about the science we did fifty years ago[3].

Another common theme in the biology reform literature is that our present system often excludes many students from pursuing science and related fields[4]. Unsurprisingly, these reports recommend that the biological community rethink the way they teach in order to provide more effective preparation for future biologists and health-care professionals and to make science more accessible and relevant in the world today[5].

### The current state of biology education reform

Responding to both of these challenges, university biology faculty have developed undergraduate biology education reforms aimed at changing the content or pedagogical structure of the curriculum. The reforms' implicit purposes are to help students develop: (i) deeper levels of understanding, (ii) transferable knowledge, and

---

[1] NRC, 1996, 1997, 1999, 2003, 2000, 2009

[2] NRC 1983, 1996, 2003, 2009
[3] NRC, 1996, 1997, 1999, 2003, 2000, 2009
[4] Tobias, 1990
[5] Handelsman et al., 2006; Hulleman & Harackiewicz, 2009; NRC, 1996, 1997, 1999, 2003

(iii) effective scientific reasoning skills[6]. A number of these goals can be achieved, at least for some students, through introducing active classroom learning activities and inquiry-based laboratories[7]. Often, such efforts reduce lecture time in favor of more interactive formats. They also urge instructors to cover less material more deeply by incorporating collaborative exercises[8]. There is a growing body of data analyzing the effectiveness of these specific strategies on improving student conceptual difficulties[9] and getting students to practice better critical thinking skills[10] in the biological classroom. Simply knowing that active learning can "work"[11] is an important first step to reform.

Reforming traditional lectures by using student-centered pedagogical approaches has led to documented student learning gains[12] in, for example, conceptual learning and argumentation skills. While both conceptual learning and improved student affect are important goals of reform, they are not sufficient. Learning to think scientifically means learning to make sense of the knowledge one is learning in terms of a basic web of principles as well as facts, and it means learning to reason in new situations flexibly and productively using those principles and facts. A student's ability to realize this and to bring to bear appropriate cognitive assets they have for learning new knowledge depends significantly on their expectations – what they bring from their previous experience with similar situations and from their interpretation of cues in the current environment that tells them "what's going on" and what is appropriate behavior.

Research in psychology and physics learning has demonstrated that expectations can play a dramatic role in how individuals perceive the situations they find themselves in and what they pay attention to in those situations. In psychology, it has been shown that expectations can cause subjects to ignore important cues in potentially life-threatening events (For example, pilots in a flight simulator ignored a plane parked in their path on a runway in order to read data projected on the windscreen)[13]. In physics[14], researchers found students failing to bring to bear knowledge and insights that they can be shown to possess as a result of their expectation that "that isn't the kind of thinking that's useful here."

It is our expectation, based on our experience in physics[15] that student expectations about the nature of the knowledge they are learning and the ways that they think it is appropriate to go about learning it will strongly affect how students approach even reformed courses. As a result, in order to better understand what barriers there are to teaching scientific reasoning and critical thinking skills, we will need to understand something about student expectations in biology. In this paper we use case studies of interviews with students to demonstrate how expectations can play a role in what students learn – and do not learn – in a reformed biology class.

## The role of expectations in the student epistemology literature

There is a growing body of research that investigates college students' epistemologies – their ideas about knowledge construction and learning[16]. The research is discipline specific, context driven, and motivated by the central idea that student ideas about knowledge construction may affect their learning[17]. These authors generally describe students' understanding about learning using one or more of three terms—attitudes, expectations, or epistemologies. In this paper we use the term "expectations" to describe a broad grouping of ideas that students might have about biology and biology learning that reflects on their in-class practice and activities. We focus our analysis to *epistemological expectations*: what students believe is the nature of the knowledge that they are learning and what it is that they should be (or are) doing in order to learn in a particular course.

Researchers have explored the influence of student expectations and epistemologies in many bodies of research such as physics[18], chemistry[19], sociology[20], psychology[21], and early and K-12 education[22]. A growing number of educational psychologists and STEM education researchers make theoretical and empirical arguments that aspects of students' epistemologies can affect learning[23]. Despite research suggesting that such student expectations may also become consequential for biology student learning, the current research literature in biology reform has not identified common expectations and approaches.

We argue that to successfully reform biology classes at the undergraduate level reformers must explore stu-

---

dents' expectations *as manifested in biology classes*. In this paper, the first in a series of papers, we begin to make a case that understanding students' expectations may help to explain the ways in which students encounter difficulties with reform and can help us to create more effective curricula the future[24]. This paper explores the range of expectations that emerged during in-depth interviews with students in an introductory biology class. These interviews illustrate how expectations can influence the ways in which students participate in and make sense of course activities.

## DESCRIPTION OF THE SETTING

All undergraduate biology majors (almost 2500 students) at the University of Maryland must complete a three-course introductory biology sequence consisting of Molecular and Cell Biology, Ecology and Evolutionary Biology, and Organismal Biology. This study focuses on this third course in the introductory sequence. Organismal Biology concentrates on the diversity, structure, and function of all organisms. Traditionally, Organismal Biology is taught in large lectures (100+ students) with little or no forms of active-engagement or discussion, such as group-work or whole class discussions.

This traditional approach toward teaching organismal biology is derided by both instructors and students alike. In such courses, the fundamental principles governing the diversity, structure, and function of all organisms do not emerge from the tsunami of isolated organismal facts.

Both due to negative feedback about the course and in response to challenges set forth by the broader scientific community for undergraduate biology education and preparation for medical education[25], the college agreed to modify the course content. The development and evaluation of the reforms was funded through the university and an NSF-CCLI grant[26].

This new "reform-minded" version of Organismal Biology focuses on highlighting the broad principles that yields a coherent unified picture of the structure, diversity, and function of organisms. While the course retains much of the subject material presented in the organismal chapters of most introductory biology textbooks[27], it emphasizes universal physical and chemical principles as well as the common genomic heritage of all life, and it and encourages students to think about organisms in terms of this organizational framework[28].

In addition to the changes in content, the instructors in one section[29] explored expanding the pedagogical tools used in order to create a more productive learning environment. In this section, the class was restructured in order to foster student engagement and reasoning as well as improved learning outcomes (as assessed on exams).

The format of this class section consists of ⅔ conventional lectures and ⅓ group active engagement (GAE) periods. While the particular pedagogical strategies varied, each GAE was designed to engage students in active inquiry practices while illustrating an important course topic. In addition to the weekly GAEs, these instructors have also added clicker questions, homework, and reading assignments to the curriculum. The exams are composed primarily of questions that require short written responses.

## METHODS

### Collection of Data

In order to understand how students' expectations manifest themselves in the context of this biology course, we draw from a multitude of sources, including: (i) faculty (field notes gathered from planning meetings as well as the instructors' verbal and written course reflections); (ii) individual students (videotaped interviews, survey data, written course evaluations, scanned copies of exams, and students' grades); and (iii) classroom data (videotaped student participation during both lecture and active engagement exercises). All of these sources have informed our analysis, but this paper specifically discusses and analyzes student expectations via illustrative vignettes selected from individual student interviews. In that data, we looked for specific interview examples in which student expectations, attitudes, etc. became salient for students. A more complete analysis will be presented in later publications.

When instructors combined content reforms with active engagement activities, group discussions and homework data we found gains in several domains including: improved student engagement, improved student perception of the course experience overall, and an improved awareness of the principles-based nature of biological knowledge[30]. These data were very promising, given the abundant literature on the efficacy of implementing active learning in the biology classroom. However, we soon discovered that our interview and exam data told a more nuanced story.

While we saw some improvements in student responses to the course (survey and interviews), the interview data indicated that individual students participated in complex and variable ways. Some students resisted the classes' shift in focus to more principles and more reasoning with physics and math. Furthermore, we inferred a possible link between students' statements regarding what they believed was necessary for success in biological learning and their actual behaviors in the course. Closer analysis of class of our interview data suggests that student expectations and ideas about what it means to learn biology are an important factor in their response to reform that had been overlooked.

---

[24] Hammer, 1989; Dweck, 2000; Redish & Hammer, 2009
[25] Handelsman et al., 2006; NRC, 1996, 1997, 1999, 2003; McCray et al., 2003; NRC, 2009; NSF, 1996
[26] NSF 09-191816
[27] Raven et al., 2000; Campbell & Reese 2008; Freeman, 2008
[28] http://umdberg.pbworks.com/w/page/8039417/FrontPage

[29] This course is co-taught by two pairs of professors each semester.
[30] Details of our findings will be provided in other papers.

From this review of interview data, we hypothesized that student perceptions, attitudes, and expectations could be one factor that influences their perception of the reforms and how they participate in the classroom.

## Selection of Data

Before discussing the illustrative vignettes, here is a brief description of how we collected the interview data, and how our definition of learning expectations emerged from this data set.

We collected approximately 40 hours of raw video data from 25 interviews with undergraduate biology students. These students were all enrolled in the reformed-oriented Organismal Biology class. We conducted interviews from the fall of 2009 through fall 2010. Most students interviewed were biology, pre-med, or pre-allied health majors. Students ranged from first-semester freshman to second-semester juniors. Even though this is the third introductory course, many students reported that this was their first biology class taken at College Park. This was usually due to their having received university credit for high school advanced placement classes, but a significant number of students also had transferred after taking classes at other universities or community colleges. We solicited interview volunteers from the class at various points during the semester. The interviews typically lasted for an hour and were video and audio recorded with the student's permission.

We initially transcribed the interviews and then coded them for instances where students talked, either explicitly or implicitly, about expectations. Using our definition of expectations, we coded all statements pertaining the nature of the knowledge that they are learning, what they felt they should be (or are) doing in order to learn, and what they felt they needed to do in order to do well in the course as expectations of learning. We paid particular attention to sections where students discussed their thoughts on the effectiveness (or ineffectiveness) of particular class activities as well as reported exam and homework strategies. We found these student reflections on their own behaviors particularly useful in understanding expectations about learning in this course.

Common themes emerged from iterative coding of utterances that provide some evidence of expectations and approaches within and across students. Using these data, we were able to document some of the expectations students have for doing well in this biology class.

From our initial corpus of 25 interviews, we identified common expectations that students expressed for this course. From these, we selected two for greater elaboration and deeper analysis: (1) students report that they expect that knowledge in this course is composed of facts gleaned from authority, and (2) students report that they expect that math and physics are not necessary in understanding lower-lever biology courses. We chose these because:

- they are explicitly about students' views regarding learning in the course,
- they appear frequently in our data corpus, and
- they have strong implications for biology reform.

After we identified our two learning expectations, we went back through our data corpus looking for specific interview examples of when students discussed, either explicitly or implicitly, the role of authority in learning biology or the value of math in the course.

## Data Analysis

This section is divided into two subsections. Each addresses a particular student expectation that we observed in the interviews. Within each subsection, we focus on several specific instances when interview participants discussed, either explicitly or implicitly, their attitudes and expectations and how those expectations influenced their reported class participation. While these represent only a fraction of the expectations the students reported, the illustrative vignettes allow us to show, with data, some of the ways in which expectations can influence student participation. In the conclusion section of the paper, we also will demonstrate that student expectations can be dynamic and susceptible to change, a result that may have implications for instruction.

**Learning Expectation I**—*The knowledge gained in this class consists of narrowly stated facts that some authority gives, and that do not need to be understood or made sense of within a larger context.*

A number of students in our interview data corpus made statements indicating that they thought the class was about particular facts, and some reported explicitly on ways that this expectation affected their behavior in the class.

*Support for Learning Expectation I*

Patrick, a sophomore biology major, reported that his introductory science classes do not reward what he called "free thinking." Patrick explained that the reason he felt restricted was he was always trying to get the "right answers." When asked if he ever used his own intuitions to think about his learning, Patrick answered: "Yes, but not in science--I always thought that-- I always felt that science is structured for a reason. It's not really a place for you to have, at least as a student, to have free thinking, because there's a penalty for free thinking, in that you get points docked, because many times you don't get the right answer." Patrick also admitted to frequently accepting a piece of knowledge as "true" simply because his professor told him it was true: "Why do I believe [that hox genes[31] evolved in the way it was described in class?]? Because two people with Ph.D.'s and 30, 45, 50 years of combined research experience told me to and I know better than to question that because I'm not that smart. I don't have that research experience and I know my limitations." Patrick explained that he has not yet acquired the level of mastery he feels is necessary for him to challenge his professors and perhaps even to understand their reasoning —he simply accepts their authority. He also reported that for most of the concepts his professors presented in this class, he could just "blindly accept it and then move on." Expanding on that thought, Patrick stated that asking him to explain how he knows that something is true involves thinking about his knowledge in a new way and is "beyond the scope" of introductory courses. Patrick's expectation that he needs to just accept knowledge from authority in introductory science courses has implications for how he approached learning, as we discuss later.

Another student, London, also reported her expectation that the professors should dispense knowledge in class. In her mind, they should explicitly provide her with facts to memorize, and then test her on those facts. London's experience with this class, however, appears to have run counter to her expectation. As a result, she found the class frustrating and confusing: "you should know this but you shouldn't memorize, and that whole idea, like, you should know but not memorize is confusing 'cause biology for me was always, like, you memorize what this does, not you understand—It's hard to explain--for me, I felt like we should know specific things—if you don't need to the exact names of all, each little protein and stuff involved, it confused me, like, if, then why are you mentioning it if we don't have to memorize it? And that's another the thing with the exams, like, why are you given so much information for the exams but you're told not to memorize all the information?"

Both Patrick's and London's expectations about learning in this class affected how they approached exams.

*Classroom implications for Learning Expectation I*

The expectation that authority is the source of knowledge affects the way students talked about how they approached exams. For example, Joe, a sophomore biology major, talked about how he reviewed old exam keys to discover what instructors focused on in student answers: "It's not how much you know, it's how you can put it into words to answer the question best. So, I may know everything, but when I see how he words the question, you have to be able to fit that mold and give him what he wants to hear. 'Cause like there'll be a question where you can give him an entire answer and it's wrong because it's missing one word that was circled on the key." Later, when reviewing a question he answered incorrectly, Joe restated that the instructors looked for a specifically worded answer. The question asked him to modify a phylogenetic tree[32] to reflect a given hypothesis. Joe drew a coherent phylogenetic tree, but not one consistent with the provided hypothesis.

Discussing his exam with the interviewer, Joe's focus on finding a right answer to satisfy the professor might have played a role in directing his attention away from analyzing the hypothetical situation requested: "[My answer] is a completely correct phylogeny, but it's just not the one they wanted. It all fits in, things evolved where they should be. But that's just not how they wanted it… I didn't understand enough to put it into their… the way they wanted." Because he expected the exam to be about singular "right" answers, it is plausible that Joseph was unable see the possibility for hypothetical questions on the exams. Joe's ideas about what the instructors valued also affected how he approached studying for later exams.

He also reported that he did not look over his earlier graded exam after receiving it back: "When I go back for the final exam, I'll look at the keys… If I got it right, it's going to be the same thing on the key. If I got it wrong, there's no point in studying it." Joe consistently stressed the importance of knowing the specific wording or example the instructors desired, leading him to look for the "right" words on the answer key.

Patrick also talked about trying to find the "key words" on exams and trying to anticipate a correct answer based on his knowledge of his professors and on his previous experiences with taking tests in this class, rather than his own knowledge of the subject matter:

> S: I tried to think about what they were looking for.
> I: What do you mean by that?
> S: I looked for key words and then tried to see how they paired up with the rest of the sentence--like the first one "haploid" and "zygote" those seem to be the main key words in that sentence and they may have thrown a concept in there, but they didn't on the last test so I didn't think they would do it on this one, so I didn't really pay attention...

---

[31] Hox genes are a highly conserved group of related genes that help determine both the basic structure and orientation (anterior/ posterior) of an organism.

[32] A phylogenetic tree is a diagrammatic representation that infers evolutionary relationships among various species.

In this segment[33], Patrick describes using the "key words" and the structure of the sentences, not his understanding of the biological phenomenon, to select the correct answer. He says he did not even really "pay attention" to the full question, and was mainly focused on figuring out how the professor structured the question. He expands on that theme:

    I: *So was that your basic test taking strategy for this? [The True/False section of the test]*
    S: *Yes [to determine the clauses]. That, and I thought every question was immediately false.*
    I: *Why was that?*
    S: *A test taking strategy that I learning from the first test.*

As he stated before, Patrick is using the format of previous tests to predict the types of questions he will encounter on future tests. Because Patrick assumes all of the statements to be false, he appears to expect that this test will be patterned like the old test. Much like with other pattern matching games, the assumption seems to be that the goal is to figure out the pattern. While it is a sophisticated test taking strategy, and one that seemed to prove successful in raising his grade for this exam, the strategy comes at the expense of him actually thinking deeply about the subject of his exam.

In contrast to Joe or Patrick, who both talked about key words, London reported that exams should only directly reflect the examples presented to her by her professors in class. In this way, she could demonstrate her knowledge of the examples by correctly reproducing them. In this example, London explains how she struggled to answer a particular question on an exam because it looked different from the version she had studied in class: "I remember this past exam--the phylogeny tree, and we definitely learned a lot about the phylogeny tree in class, so I was, like ok, I studied this except it was entirely backwards, like, almost a mirror image, but not quite--I was, like, crap (laughs) like, I know you're supposed to memorize it frontward and backward and, like, understand how it branches but, it was just an unexpected, like, thing [to see on the exam]." London expected to have to reproduce the tree from class, and understand how that tree branched out, not to have to construct a similar but different tree from her understanding of phylogenetic trees. To London, this failure to understand the way the knowledge in the class was structured and tested appears to follow naturally from her stated expectations about the nature of knowledge in the class.

In another example, London explains that the exam asked her to describe and use the principles learned in class to compare the gill structures of several organisms. London explained that, while she remembered the "professor mentioning the concepts briefly in class," she did not recall the professor discussing the specific organisms used on the exam. The idea that an exam question would require her to apply the general concept of gill structures to unfamiliar organisms "hit her out of nowhere." Once again, her description is completely consistent with her expectation that exam questions should come directly from lecture material and not require independent thought.

**Learning expectation II**—*The role of math and physics in biology is useful at some level, but not necessary (or even appropriate) at the level of the class they are in (a second year college biology class).*

*Support for Learning Expectation II*

In interviews, some students questioned the value of using equations to learn biology. For example, Judy reported that she perceived that math did not help her learn biology: "I don't see how, other than, you know, statistics, usually it doesn't seem like math really comes into play a lot of times you know--Maybe I'm wrong and it does come into play a lot, but other than things like the physics of flow in your heart, and I could care less about the physics that go on in your heart because I'm not going to become a doctor, and as far as statistics umm, I am going to take a biostats course for that."

Judy had difficulty seeing how math is important in learning biology. For her, it seems like math rarely "comes into play" when she thinks about biological phenomenon. Even in her example about the heart, Judy explains that the math might be helpful to some, but not to her, because she does not want to be a doctor. Overall, she does see some math, like statistics, as relevant to her future endeavors, but she does not believe that statistics need be incorporated into her biology classes in order for it to be useful to her understanding biology.

Given that this course was specifically designed to teach physical, chemical, and mathematical principles in order to help students understand biology, it is interesting that Judy believes that she can learn the "principles" and "concepts" of this course without physics or math. One possibility is that her expectation that math and physics is not relevant to biology understanding undermines her ability to pay attention to what the professor in the class is trying to show—that math helps one understand many things in biology, including the relation of structure and scale. Another is that she is picking up on cues that the class could be (inadvertently) sending that although the math and physics are relevant in principle, that in practice it can be ignored.

---

[33] The original exam question was written: Frequently, the haploid zygote is resistant to adverse environmental conditions. (True or False. If false, rewrite statement).

Hall et al.　　　　　　　　　　　　　　　　　　　　　　　　　　　　The Impact of Student ExpectationsheaderHall et al.　The Impact of Student Expectations

*Classroom implications for Learning Expectation II*

Students' views of the value of math and physics influenced how they participated in specific physics-based GAEs. For example, after her instructor used Fick's laws[34] of diffusion in a series of lectures primarily as a referent in understanding the affordances and constraints in evolutionary development, one student, Ashlyn, questioned the approach: "I think that biology is just—it's supposed to be tangible, perceivable, and to put that in terms of letters and variables is just very unappealing to me, because like I said, I think of it as it would happen in real life, like if you had a thick membrane and you try to put something through it, the thicker it is, obviously the slower it's gonna go through. But if you want me to think of it as 'this is $x$' and 'that's $d$' and then 'this is $t$,' I can't do it. Like, it's just very unappealing to me."

During the interview, Ashlyn told us that she resisted using letters and symbols when discussing the equations in a specific activity about diffusion. One reason Ashlyn found using the equations "unappealing" and of little general use in her diffusion exercise is because she already understood the concept beforehand. She explained that before going into the exercise, she already knew "how diffusion worked" and could describe the phenomenon qualitatively in terms of membrane thickness and molecule size. She went on to explain that most biological situations, like diffusion, only require this qualitative or descriptive understanding of the phenomenon: "So the equation, like I said before, like, I will memorize it because I have to, but knowing that, it's-- the time is directly proportional to distance and indirectly proportional to the diffusion constant, I think in my mind is enough."[35] Due to her views about the value of equations, Ashlyn admitted to "blocking" the equations from her general understanding biology and only memorized them in order to be able to do the specific calculations she thought would be on the class exams.

Another reason Ashlyn resisted using equations to understand diffusion is because she did not see how the equations improved her ability to express the concepts: "it's basically a way to put it, put the concept into words. I think that's what the only function of the equations are." She finds thinking about diffusion in terms of equations as being just another way to verbalize concepts. Ashlyn understands that, in principal, equations express concepts, but doesn't see the value of doing so. Moreover, she feels that the approach undermines what attracts her to biology. Even after the exercise, she still does not like thinking about biology in terms of letters and variables. Because Ashlyn did not expect equations to add to her understanding of diffusion, she did not feel like she took much away from this particular activity. Perhaps if Ashlyn had a different perception about what equations could do for her in this context, she would have been able to find more value in them and in this particular exercise.

Just as the students' views of math and physics affected their class participation, they also affected their approaches to exam problems. On exams, Joseph sees the utility of equations as limited because he assumes that biology professors cannot expect their students to do much with them. When asked if he saw the small amount of math as helpful, Joseph answered "no" because the math that was in the course was too simple to really explain anything to him. Following that statement, Joseph reported that, on the positive side, he was able to earn some extra exam points with some easy calculations, but he still did not see the math as providing him with a deeper, more complex understanding of the information:

> I: So is it useful to have the math here (to understand the limits of diffusion in the context of this exam question?)?
> S: I mean it's an easy problem. An easy three points.
> I: Did that help your understanding of biology?
> S: No. I mean this is such a simple concept. It was just extra work [on the exam].

He appears just focused on doing the calculations, and not on what the calculations could do for him.

## IMPLICATIONS

In addition to identifying some specific expectations students had about what they were learning in the class, our data indicated two additional things that have implications for instruction. First, students did not all progress towards more sophisticated views about the nature of biological knowledge and biology learning. Second, we found that some students may not automatically be "on board," with the implicit goals set out for them by their instructors, and instead resist reform by maintaining their own expectations about learning biology.

### Expectations can be dynamic – students can shift

To illustrate the first point, we return to our discussion of Ashlyn. For most of her interview, her reported views about the class appear directly oppositional to the stated goals of reform, in particular, to the explicit use of math and physics as organizing and explanatory tools. However, at one specific point in the interview, Ashlyn reports a different perception about the value of using mathematics and physics to explain biological phenomenon. She responded in a strong positive way to using

---

[34] Fick's two laws predict (i) the direction and speed of diffusion and (ii) how diffusion causes the concentration to change with time.

[35] This is, of course, one of the problems of not using the math. The time is *not* directly proportional to the distance but to the square root of the distance. This is one thing that the equation, or even just the units of the symbols involved, would tell you.

mathematics when discussing a *specific course activity about scaling*.

She recalled a demonstration in which two wooden horses were held next to each other, one of which was twice the size of the other (each dimension was scaled by 2). The small horse was able to stand, while the larger horse collapsed. In subsequent discussions, the students and instructor worked through the mathematical relationship between surface area and volume. Ashlyn stated that she found this exercise particularly helpful for understanding biology: "The little one and the big one, I never actually fully understood why that was. I mean, I remember watching a Bill Nye episode about that, like they built a big model of an ant and it couldn't even stand. But, I mean, visually I knew that it doesn't work when you make little things big, but I never had anyone explain to me that there's a mathematical relationship between that, and that was really helpful to just my general understanding of the world. It was, like, mindboggling."

Although she talked about this demonstration in the same interview where she spoke of the unappealing nature of equations, she voiced a very different opinion about the usefulness of mathematics in understanding biology, now finding it "really helpful" and "mindboggling" rather than "unappealing" and pointless. Not only does Ashlyn talk differently about the value she sees in these two examples, her entire demeanor appeared to change. When describing the diffusion activity, she appeared rigid and frowning. By contrast, she was smiling, leaning forward and talking rapidly and excitedly when discussing the activity on scaling. These quotes show that students' can shift the way they interpret and respond to the class based on the content, instructional environment, or other contextual cues.

In order to maximize the benefits of reforms, it seems appropriate that instructors be aware of and explicitly and consistently address the multitude of students' expectations, attitudes, beliefs etc. about what biology is, what it means to learn and understand in biology, and the ways that this course will help to achieve those goals.

## Expectations can be "sticky" and get in the way

Our interview data also indicate that some students have robust expectations about learning that impede the successful implementation of even well orchestrated reforms. For example, students may reject the reforms as unhelpful and pointless, as seen in Ashlyn's diffusion example, and decide not to participate in the exercise at all, or they may misinterpret their role as students in the learning process. In this example, Ashlyn explains how she has learned to diligently and methodically memorize lectures in this course, even the "what do you call it? The… concepts and generalizations." Clearly, this is not what the professors of the course had intended their students to take away from the class!

An instructor may change the (content and pedagogical) focus in her class to a principles-based course, but that does not guarantees that her students will modify their expectations about how to do well on her exams or approach learning. In order for reforms to succeed, it is important to also consider the students' expectations, goals, and objectives, independent of those set out by the course and the instructor, and to realize that just telling students that the situation has changed may not suffice to get them to change inappropriate in-class attitudes and behaviors. Meta-messages left over from extensions of traditional pedagogy, statements interpreted on way by a faculty member and another way by students, and even "the word on the street and the internet" about the class from previous students can inadvertently confirm students' inappropriate expectations. When such misalignments go ignored and unaddressed in the classroom, it may undermine even carefully orchestrated reforms.

## Discussion and Summary

A growing movement among biology educators has urged a rethinking of introductory biology courses in order to both address problems within the current system and to foster more sophisticated ways of thinking about biology and effective scientific reasoning skills. To accomplish these goals, most efforts so far focus on content and pedagogy: instructors are urged to "get over coverage" and, instead, concentrate on incorporating collaborative active learning strategies and other reformed pedagogical approaches in order to emphasize thinking over memorization[36]. We agree and these recommendations served as the starting place for our own course reforms. However, previous research on curricular change and our own data now suggest, that many students may not benefit from these changed courses unless the reforms also take into account —and try to change — students' epistemologies and expectations. By this we mean their views about what counts as knowing and understanding in biology and about what kinds of knowledge and learning specific courses reward[37].

---

[36] Handelsman et al., 2006; Hulleman & Harackiewicz, 2009; NRC, 1996, 1997, 1999, 2003

[37] NRC, 1996; 1997; 1998; 1999; 2003; 2009

Hall et al.                                                                                                    The Impact of Student Expectationsment for measuring student beliefs about physics and learning physics: The Colorado Learning Attitudes about Science Survey. *Physical Review Special Topics - Physics Education Research,* 2(010101).

Allen, D. E., & Duch, B. J. (1998). *Thinking Toward Solutions: Problem-based learning activities for general biology.* Harcourt College Publishing.

Belenky, M. F., Clinchy, B. M., Goldberger, N. R., & Tarule, J. M. (1986). *Women's Ways of Knowing.* Basic Books.

Blumberg, P., & Michael, J. A. (1992). Development of self-directed learning behaviors in a partially teacher-directed problem-based learning curriculum. *Teaching and learning in medicine,* 4(1), 3–8.

Blumenfeld, P. C., Soloway, E., Marx, R. W., Krajcik, J. S., Guzdial, M., & Palincsar, A. (1991). Motivating project-based learning: Sustaining the doing, supporting the learning. *Educational Psychologist,* 26(3), 369–398.

Boyes, E., & Stanisstreet, M. (1991). Misconceptions in First-Year Undergraduate Science Students about Energy Sources for Living Organisms. *Journal of Biological Education, 25*(3), 209-213.

Buehl, M. M., & Alexander, P. A. (2006). Examining the dual nature of epistemological beliefs. *International Journal of Educational Research,* 45(1-2), 28–42.

Campbell, N. A., & Reece, J. B. (2008). *Biology Eighth Edition (8th ed.).* Benjamin Cummings.

Conley, A., & Pintrich, P. (2004). Changes in epistemological beliefs in elementary science students. *Contemporary Educational Psychology,* 29(2), 186-204.

Dweck, C. S. (2000). *Self-theories: Their role in motivation, personality, and development.* Psychology Press.

Ebert-May, D., Brewer, C., & Allred, S. (1997). Innovation in large lectures: Teaching for active learning. *Bioscience,* 47(9), 601–607.

Eisobu, G. O., & Soyibo, K. (1995). Effects of concept and vee mapping under three learning models on students' cognitive achievement in ecology and genetics. *Journal of Research in Science Teaching,* 32, 971–995.

Elby, A., & Hammer, D. (2001). On the substance of a sophisticated epistemology. *Science Education,* 85(5), 554–567.

Freeman, S. (2008). *Biological Science (3rd ed.).* Benjamin Cummings.

Garvin-Doxas, K., & Klymkowsky, M. (2008). Understanding randomness and its impact on student learning: lessons learned from building the Biology Concept Inventory (BCI). *CBE-Life Sciences Education,* 7(2), 227-233.

Garvin-Doxas, K., Klymkowsky, M., & Elrod, S. (2007). Building, using, and maximizing the impact of concept inventories in the biological sciences: report on a National Science Foundation sponsored conference on the construction of concept inventories in the biological sciences. *CBE-Life Sciences Education,* 6(4), 277-282.

Goodwin, H. A., & Davis, B. L. (2005). Teaching undergraduates at the interface of chemistry and biology: challenges and opportunities. *Nature Chemical Biology,* 1, 176-179.

Grove, N., & Bretz, S. (2007). CHEMX: An Instrument To Assess Students' Cognitive Expectations for Learning Chemistry. *Journal of Chemical Education,* 84(9), 1524-1529.

Hammer, D. (1989). Two approaches to learning physics. *The Physics Teacher,* 27(9), 664–670.

Hammer, D., & Elby, A. (2002). On the form of a personal epistemology (in B. K. Hofer & P. R. Pintrich eds.). *Personal epistemology: The psychology of beliefs about knowledge and knowing,* 169–190.

Handelsman, J., Miller, S., & Pfund, C. (2006). *Scientific Teaching.* WH Freeman & Co.

Hofer, B. K. (2000). Dimensionality and disciplinary differences in personal epistemology. *Contemporary Educational Psychology,* 25(4), 378–405.

Hofer, B. K. (2001). Personal epistemology research: Implications for learning and teaching. *Educational Psychology Review,* 13(4), 353-383.

Hofer, B. K. (2002b). Personal epistemology as a psychological and educational construct: An introduction. In B. K. Hofer & P. R. Pintrich (Eds.), *Personal epistemology: The psychology of beliefs about knowledge and knowing* (p. 3–14). Mahwah, NJ: Lawrence Erlbaum Associates, Inc.

Hofer, B. K. (2006). Domain specificity of personal epistemology: Resolved questions, persistent issues, new models. *International Journal of Educational Research,* 45(1-2), 85–95.

Hofer, B. K., & Pintrich, P. R. (1997). The development of epistemological theories: Beliefs about knowledge and knowing and their relation to learning. *Review of Educational Research,* 67(1), 88-140.

B. K. Hofer & P. R. Pintrich (2002). *Personal Epistemology: The psychology of beliefs about knowledge and knowing.* Mahwah, NJ: Lawrence Erlbaum Associates, Inc.

Hulleman, C. S., & Harackiewicz, J. M. (2009). Promoting interest and performance in high school science classes. *Science,* 326(5958), 1410.

Khodor, J., Halme, D. G., & Walker, G. C. (2004). A hierarchical biology concept framework: a tool for course design. *CBE-Life Sciences Education,* 3(2), 111-121.

Limón, M. (2006). The domain generality–specificity of epistemological beliefs: A theoretical problem, a methodological problem or both? *International Journal of Educational Research,* 45(1-2), 7-27. doi: 10.1016/j.ijer.2006.08.002.

Lising, L., & Elby, A. (2005). The impact of epistemology on learning: A case study from introductory physics. *American Journal of Physics,* 73(4), 372.

McCray, R. A., Dehaan, R. L., & Schuck, J. A. (2003). *Improving Undergraduate Instruction in Science, Technology, Engineering, and Mathematics: Report of a Workshop.* Washington, D.C. The National Academies Press.

Michael, J. (2006). Where's the evidence that active learning works? *Advances in physiology education,* 30(4), 159-167.

Michael, J., Wenderoth, M. P., Modell, H. I., Cliff, W., Horwitz, B., McHale, P., et al. (2003). Undergraduates' understanding of cardiovascular phenomena. *Advan. Physiol. Edu.,* 26(2), 72-84.

Most, S., B. Scholl, B., Clifford, E., & Simons, D. (2005). What you see is what you set: Sustained inattentional blindness and the capture of awareness. *Psych. Rev.,* 112(1), 217-242.

National Commission on Excellence in Education. (1983). *A Nation at Risk: The imperative for educational reform.* Washington, D.C. Retrieved from http://www2.ed.gov/pubs/NatAtRisk/index.html.

National Research Council. (1996). *From Analysis to Action: Undergraduate education in science, mathematics, engi-
*AERA 2011*                                                                                                                                                                9